\documentclass{optica-article}

\journal{opticajournal} 

\articletype{Research Article}

\usepackage{lineno}

\usepackage{graphicx}
\usepackage{dcolumn}
\usepackage{bm}
\usepackage{hyperref}

\usepackage{makecell}

\usepackage{subfig}
\usepackage{subcaption}

\usepackage{xcolor}
\usepackage{soul}
\usepackage{easyReview}

\begin{document}

\title{Quantifying the advantage of quantum correlation microscopy using arrays of single-photon detectors}

\author{Jaret J. Vasquez-Lozano,\authormark{,1*} Qiang Sun,\authormark{1,†} Shuo Li\authormark{1,‡} and Andrew D. Greentree\authormark{1,§}}

\address{\authormark{1 ARC Centre of Excellence for Nanoscale BioPhotonics, School of Science, RMIT University, Melbourne 3001, Australia}}

\email{\authormark{*}jaret.vaslo@gmail.com}
\email{\authormark{†}qiang.sun@rmit.edu.au}
\email{\authormark{‡}shuo.li.rmit@gmail.com}
\email{\authormark{§}andrew.greentree@rmit.edu.au}



\begin{abstract*} 
Quantum correlation microscopy is an emerging technique for improving optical resolution.  By taking advantage of the quantum statistics from single-photon fluorophores, more information about the emitters (including number and location) is obtained compared with classical microscopy.  Although it is known that the resolution can be improved by increasing detector numbers, as well as using quantum correlation, the quantitative relationship between these two approaches is not immediately clear. Here we explore widefield quantum correlation microscopy using arrays of single-photon detectors. We explicitly compare the use of $N$ detectors used in photon counting mode vs $N/2$ detectors used to measure quantum correlations. i.e., where there are $N/2$ Hanbury Brown and Twiss systems, using the same $N$ detectors, on randomly generated two-emitter systems. We find regimes where $N/2$ Hanbury Brown and Twiss detectors provide improved localisation compared to $N$ photon counting detectors, as a function of emitter position and number of photons sampled.
\end{abstract*}

\section{Introduction}

The optical diffraction limit imposes a restriction on the ability to resolve light-emitting particles that are close with respect to the wavelength of light \cite{Rayleigh,Abbe}. This limit can be overcome through the use of additional resources such as \emph{a priori} knowledge, nonlinear optics, or similar.  This has led to the development of new super-resolution techniques, such as Stimulated Emission Depletion (STED) and Stochastic Optical Reconstruction Microscopy (STORM) \cite{Hemmer16,Klar,Rust,Hemmer12,Sauer,Tam,Diaspro}. However these techniques all come with the restriction of increased imaging intensity, longer acquisition times, or both. 

Quantum correlation microscopy (QCM) is a diffraction unlimited imaging technique where the quantum properties of emitters are used to improve the ability to localise and quantify the number of emitters without the need of increased intensity or aquisition time. The original proposal is due to Hell, Soukka and H\"{a}nninen \cite{Hell} who show that use of higher-order Hanbury Brown and Twiss correlations reduces the optical point spread function by a factor of $1/\sqrt{k}$ where $k$ is the order of photon correlations.

Recent works \cite{Gatto,Schwartz12,Schwartz13,Classen,Bartel,Tenne,Worboys,Shuo,Davin,Jaret} have all demonstrated the ability of QCM to improve resolution when imaging single-photon emitting sources such as those found in diamond color-centers \cite{Eisaman} by taking advantage of photon anti-bunching.

Single photon avalanche detectors (SPAD) are an important and rapidly developing technology with wide applicability to microscopy\cite{Rochas,Zappa,Bronzi}.  In particular, SPAD arrays are now being used as a tool for wide-field microscopy and photon-emitter localisation \cite{Bruschini,Arin}.  SPAD arrays are typically used in a conventional mode, where each SPAD acts as a pixel for wide-field imaging.  However, it is also possible to use the detectors in a correlated mode, where the outputs from detectors are correlated to operate as Hanbury Brown and Twiss detectors \cite{Antolovic,Israel}.  This is seen in, for example, Boiko \textit{et al.} \cite{Boiko} where they design a a SPAD array system to perform QCM.

Here we show a direct numerical comparison between the localisation of two particles from $N$ SPAD detectors in a square array, used in conventional (independent) mode, and $N/2$ detectors in Hanbury Brown and Twiss (HBT) mode, where pairs of detectors are registered to the same region of interest through a beam splitter and are used in correlation.  Our results show that in many cases the $N/2$ detectors provide superior localisation than the $N$ detectors, with the actual advantage depending on the precise configuration of the emitters relative to the grid.  Independent operation is typically superior when the emitters are well localised, and when they are far enough apart that the correlation signal is low compared to the direct detection.

We first compare the time-scaling results for intensity-only and HBT for our simulated widefield system. We then show the scaling results as a function of detector spacing at fixed measurement times. We compare the scaling of intensity-only and HBT for specific numbers of detectors to show in what regimes the use of second-order correlations is justified for emitter localisation. Finally, we demonstrate that the advantage gained from HBT increases as a function of photon detection efficiency.

\section{Theory} 

\subsection{Optical point spread function and photon counting} \label{sect:2A}

\begin{figure}[tb!]
    \centering
    \makebox[\linewidth]{
    \includegraphics[width = 1.2\linewidth]{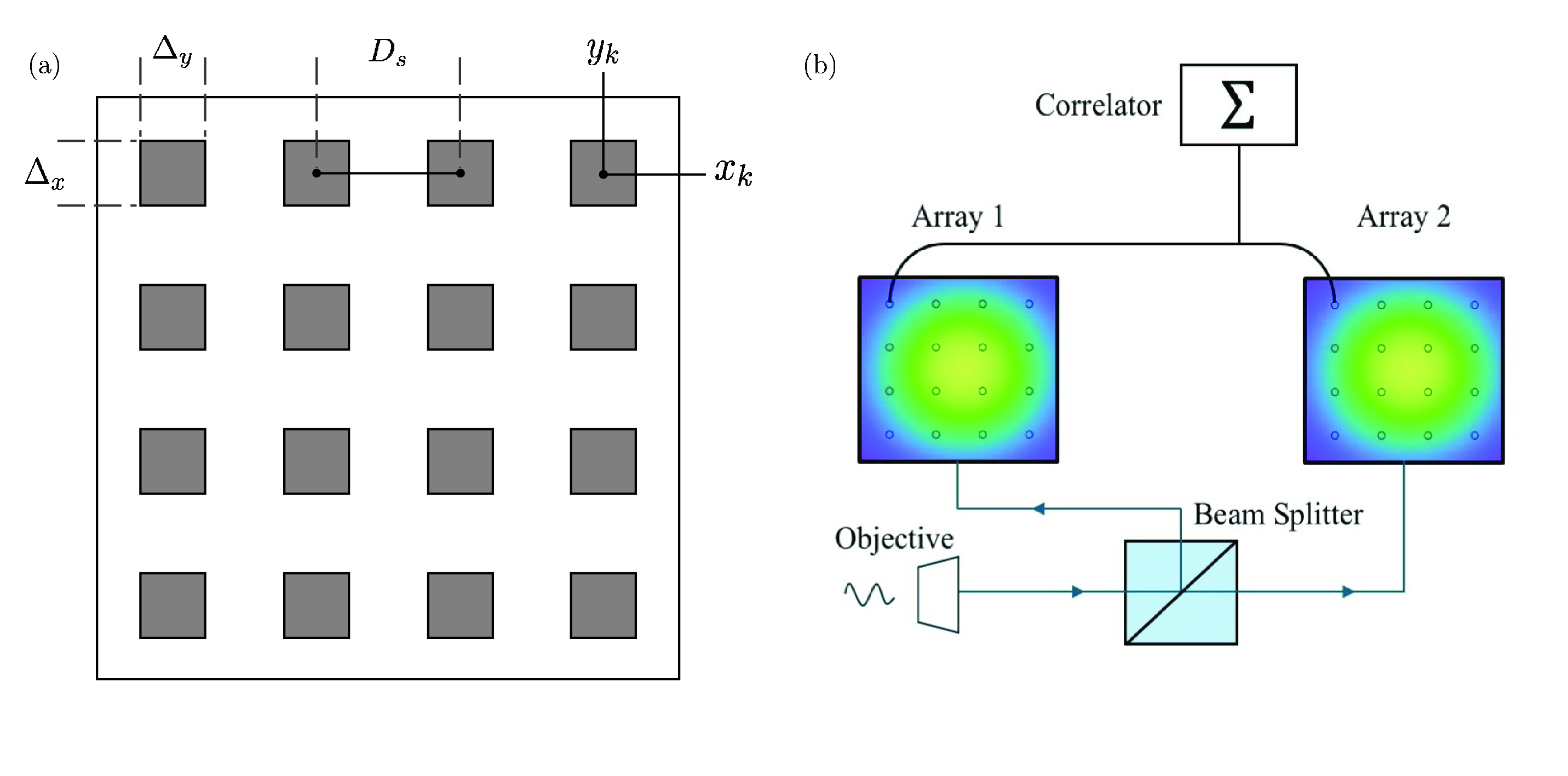}}
    \caption{(a) Schematic of example SPAD array of $N = 16$ detectors. The gray squares represent the active detection area of each SPAD chip. $\Delta_x$ \& $\Delta_y$ are the length of the sides of the active area. We treat the active area of the SPAD chip as the entire SPAD for simplicity. $x_k$ \& $y_k$ are the center coordinates of each SPAD. $D_S$ is the distance from the center point between each adjacent SPAD. (b) Schematic of HBT setup with twin SPAD arrays. The light signal is split via beam splitter to two identical arrays. Individual SPADs are marked as circles. Photon correlations are performed between detector pairs, where each detector in the pair monitors the same region of the image plane, as shown by connection to correlator.}
    \label{fig:SPAD_Schem}
\end{figure}

To simulate photon detections and photon correlations for an experiment, we begin by modeling the optical point spread function (PSF) of our emitters. We imagine a system comprised of $N$ equally spaced SPAD detectors arranged in a square grid. An example schematic for $N=16$ can be seen in Figure~\ref{fig:SPAD_Schem}(a).

We estimate the probability of photon detection for a given detector by integrating a Gaussian PSF, which is a sufficient approximation to the Airy disk PSF \cite{Stallinga}, over the detector area:

\begin{align}
    P_{i,k} = \int^{y_k + \Delta_y /2}_{y_k - \Delta_y /2} \int^{x_k + \Delta_x /2}_{x_k - \Delta_x /2} (1 - \eta) \frac{P_{B,i}}{2\pi \sigma^2} \exp\left[ -\left(\frac{(x-x_{0,i})^2 + (y-y_{0,i})^2}{2\sigma^2}\right)\right],
\label{eq:P_i INTEG}
\end{align}
where $x_k$ and $y_k$ are the center coordinates of the SPAD detector, $k$. $\Delta_x$ \& $\Delta_y$ are the lengths of the SPAD's active detection region along the $x$ and $y$ axis, and we assume that the active detection region of each SPAD is square, such that $\Delta_x = \Delta_y$. $x_{0,i}$ \& $y_{0,i}$ are the center coordinates of the emitter PSF, i.e., the ground truth position of emitter $i$. The parameter $\eta$ takes account of the optical losses in the system. $P_{B,i}$ is the intrinsic brightness of the emitter. $\sigma$ is the standard deviation of the PSF and
\begin{align}
    \sigma \approx \frac{0.21 \lambda}{\text{NA}},
\label{eq:sig}
\end{align}
where $\lambda$ is the wavelength, and NA the numerical aperture. For generality, we express all distances in units of $\sigma$.

The full solution to Eqn.~\ref{eq:P_i INTEG} is: 
\begin{gather}
    P_{i,k} = (1 - \eta)  \frac{P_{B,i}}{4} \left[ \text{erf}\left(\frac{2y_k +\Delta-2y_{0,i}}{2\sqrt{2}\sigma} \right)+\text{erf}\left(\frac{-2y_k +\Delta+2y_{0,i}}{2\sqrt{2}\sigma}\right)\right]\nonumber\\ \times \left[ \text{erfi}\left(\frac{x_k-\frac{\Delta}{2}-x_{0,i}}{\sqrt{2}\sigma}\right)-\text{erfi}\left(\frac{x_k+\frac{\Delta}{2}-x_{0,i}}{\sqrt{2}\sigma}\right)\right].
\label{eq:P_Erf}
\end{gather}

Expanding in powers of $\Delta$, the first non-zero correction is:

\begin{align}
    P_{i,k} = (1 - \eta)  \frac{P_{B,i}\Delta ^{2}}{2\pi \sigma^2} \exp\left[ -\left(\frac{(x-x_{0,i})^2 + (y-y_{0,i})^2}{2\sigma^2}\right)\right],
\label{eq:P_i}
\end{align}
\noindent
which we use the calculations below.

In our simulation, each pulse cycle is a photon sampling experiment. We envision a pulse-laser excitation system, where single-photon emitters are excited by a pulse that is short compared to the emission time. The emitted photons are detected by one of several SPADs during a collection window that is large relative to the florescence lifetime of the emitters. This process is repeated over several cycles where the timing between pulses is large relative to the dead-time of the SPAD. Here, the number of experiments is analogous to the integration time of the measurement of the sample, which would be the number of pulse cycles multiplied by real duration of the experiment (i.e, $10^8$ pulse cycles or experiments with a duration of 100 ns each would be equivalent to an integration time of 10 s.).

To simulate photon detections for this system, we obtain a random number of counts, $c_i$, from a normal distribution using MATLAB's normrnd function for each measurement location. The expected value, $\mu$ and standard deviation $s$, are based on the expected number of detections and standard deviation for a multinomial distribution for probabilities $P_i$ for $n$ number of experiments. We approximate the multinational distribution with a normal distribution to bypass computational restrictions of having large $n$ values. Due to $n$ being large, any positive bias effects are negligible. In any instances of $c_i <0$, we set $c_i$ to 0.
\begin{align}
\mu &= n P_i, \nonumber \\
s & = \sqrt{n P_i (1-P_i)},  \nonumber \\
c_i &= \text{normrnd}(\mu,s),
\label{eq:c_i}
\end{align}
where $\text{normrnd}$ is the normal random number generator of Matlab \cite{Mat}.

\subsection{Photon Correlations}

Photon correlation measurements are typically obtained via the Hanbury Brown and Twiss experiment (HBT). An example schematic is shown in Figure~\ref{fig:SPAD_Schem}(b). The intensity signal is split to two detectors using a beamsplitter, and the coincidence rate of photons is obtained by monitoring the detection delay times between detectors. HBT detectors are capable of measuring correlations and intensity by summing the detections at each detector. The use of HBT in this work refers to the use of both correlated photon counts and individual photon counts, i.e., intensity.

In order to simulate the HBT results for our system, we consider two emitters, $i$ and $j$. We use the photon detection probabilities of each emitter, $P_i$ and $P_j$, to calculate the expected number of co-incident photon detections, $c_{i,j}$. Randomness due to photon fluctuations is introduced by once again using MATLAB's normrnd function: 
\begin{align}
\mu_c &= \frac{1}{2}nP_i P_j ,   \nonumber \\
 s_c &= \sqrt{\frac{1}{2}n P_i P_j \left(1-\frac{1}{2} P_i P_j\right)} ,      \nonumber \\
c_{i,j} &= \text{normrnd}(\mu_c, s_c).
\label{eq:c_ij}
\end{align}
\noindent
where $n$ is the number of experiments, $\mu_c$ is the expected value of the coincident counts for a multinomial distribution, and $s_c$ is the standard deviation of the multinomial distribution. In our scenario, on average, half the photon pairs will be sent to the same detector due to the beam splitting ratio, resulting in no photon coincidence detection. To account for this, we introduce a $1/2$ factor here.

For our simulated widefield HBT system, incoming light is sent to two identical SPAD arrays, where photon correlations are measured for detector pairs with the same relative position in their SPAD array, as shown in Fig.~\ref{fig:SPAD_Schem}(b). We account for any optical losses by $\eta$ as shown in Eqn.~\ref{eq:P_i}.

The second order photon correlation function may be written as a ratio of correlated to uncorrelated photons, i.e., where correlations at detection delay time $\tau=0$ is  divided by correlations where $\tau = \pm\infty$. For $E$ emitters, the second order correlation function is \cite{Worboys}:
\begin{equation}
    g^{(2)}_E = \frac{2\sum_{i=1}^{E-1}\sum_{j=i+1}^{E}P_i P_j}{\sum_{i=1}^{E}\sum_{j=1}^{E}P_i P_j},
\label{eq:g2N}
\end{equation}
We choose to use only coincident photon counts as opposed to $g^{(2)}$ to keep intensity and correlation data in units of counts, which simplifies the Least Squares Estimator we use in our fitting.

\subsection{Effective point spread function}

The equations introduced thus far are generalized for any number of emitters. In this work, we will only consider the case of two emitters.

To obtain the necessary statistics to determine scaling, we perform 200 Monte-Carlo simulations for every each random emitter configuration for a given number of experiments. Emitter 1's brightness, $P_{B,1}$, is fixed, while emitter 2's brightness, $P_{B,2}$, is randomized. We randomize emitter spacing, $d$, with position bounds at $x = (-0.2,0.2)$ and $y = (-0.2,0.2)$.

By minimizing a Least Squares Estimator, we optimize a fit to our random simulated data to obtain a localisation estimate for each trial. 

\begin{gather}
    \text{RSS} = \sum^{N}_{k = 1} (1-C_{W,k})(I_{\text{obs.},k}-I_{\text{expec.},k})^2 + C_{W,k}(C_{\text{obs.},k}-C_{\text{expec.},k})^2, \nonumber \\
    C_{W,k} = \frac{\sum^{N}_{k = 1}C_{\text{obs.},k}}{\sum^{N}_{k = 1}C_{\text{obs.},k}+I_{\text{obs.},k}},
\label{eq:LSE}
\end{gather}
\noindent
where $N$ is the total number of detectors, $k$ is the detector index, $I$ is the expected or observed photon counts at the detector, and $C$ is the expected or observed coincident counts at the detector. $I$ and $C$ is normalised for fitting, so an effective weighting, $C_W$, is used to preserve the difference in relative values of photon counts and coincident counts \cite{Jaret}. When considering intensity-only detectors, $C_W$ is set to zero.

Following the approach in \cite{Worboys} we create a metric for localisation uncertainty by creating a polygon containing the smallest cluster with 39.5\% of the localisation estimates, i.e., containing one standard deviation worth of points for a two-dimensional Gaussian. From this polygon, we create an effective PSF, $w_{\text{eff}}$ that is comparable to resolution width in conventional microscopy.
\begin{equation}
    w_\text{eff} = 2\sqrt{\frac{A_{39.5\%}}{\pi}}
\label{eq:Weff},
\end{equation}
where $A_{39.5\%}$ is the area of the smallest polygon that contains $39.5\%$ of the data.

\begin{figure}[tb!]
    \centering
    \includegraphics[width = 0.7\linewidth]{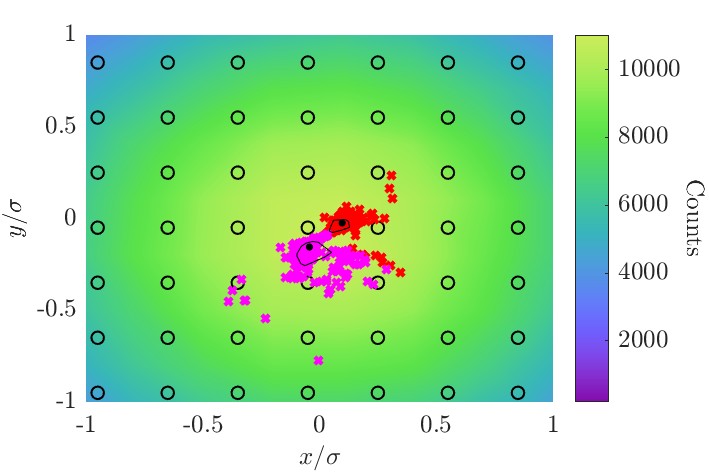}
    \caption{Example collection of localisation estimates for two emitters, shown in red and magenta. The data simulates 200 Monte-Carlo trials for 2$\times 10^{7}$ photon sampling experiments, with emitters spaced $0.192\sigma$ apart and brightness' of $P_{\eta,1} = 0.4$ \& $P_{\eta,2} = 0.364$. The resulting $w_{\text{eff}}$ for each emitter is shown by black contours. Black dots are ground truth locations. Hollow dots are detector locations.}
    \label{fig:Weff}
\end{figure}

An example configuration using this process is shown in Fig.~\ref{fig:Weff}. As each two-emitter configuration results in two individual $w_{\text{eff}}$ values (one for each emitter), we simplify our results by taking the average of our two $w_{\text{eff}}$. We denote this value as $\overline{w_{\text{eff}}}$.

\section{Results}

\subsection{localisation precision with respect to number of experiments for random emitter configurations}\label{sect:TimeLoc}

From the work shown in \cite{Jaret}, we see that an increase in the number of photon coincidences, either through the reduction of space between emitters or the introduction of background emitters, improves the localisation precision obtained using HBT plus intensity. This is reflected in the time scaling results of the work, which show how the time required to localise the emitters using HBT compared to intensity-only detectors is much lower when the number of coincident photons is increased.

To see how the localisation precision of our widefield HBT system scales with time, or number of experiments in our case, we calculate $w_{\text{eff}}$ as a function of the number of experiments, $n$. For the system described in Section~\ref{sect:2A} and Eqn.~\ref{eq:P_i}, we set $\Delta^2$ to $1/200$. This ensures that when reducing detector spacings up to $0.05\sigma$, that $\sum^{N}_{k=1}P_{i,k} < 1$ at the limit where $\eta = 0$. For simplicity, we combine $P_{B,i}$ and $\eta$ such that 

\begin{equation}
    (1-\eta) P_{B,i} =P_{\eta,i}
\label{eq:P_eta}.
\end{equation}

We leave the default value of $P_{\eta,1}$ as 0.4. The relative value of $P_{\eta,2}$ will range from $P_{\eta,1}$ to $0.5P_{\eta,1}$. The work shown in \cite{Worboys} notes a change in scaling behavior when the relative brightness is less than 0.5, hence the lower bound.

\begin{figure}[tb!]
    \centering
    \includegraphics[width = 0.7\linewidth]{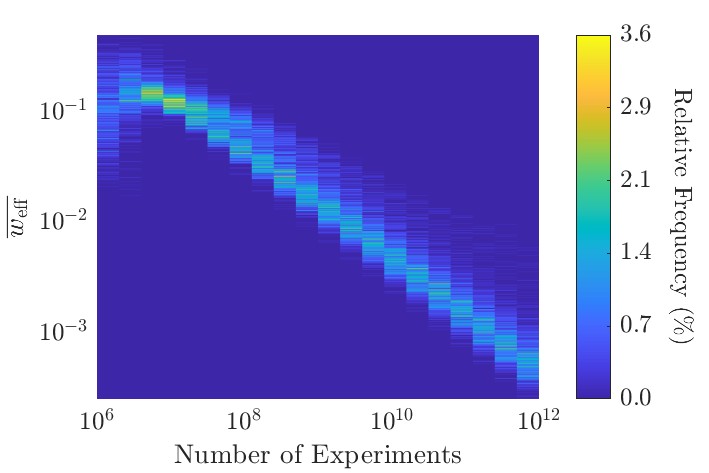}
    \caption{$\overline{w_{\text{eff}}}$ (mean of two emitter's $w_{\text{eff}}$ values per case) as a function of number of experiments, $n$, for 700 unique emitter cases with emitter spacings randomly selected up to a maximum of $d = 0.57\sigma$. After a certain number of experiments for each configuration ($n_{\text{tp}}$), the majority achieve a scaling of $1/\sqrt{n}$. The number of experiments required to achieve this scaling is dependent on the spacing of the emitters, with a higher number required for smaller spacings. Localisation prior to $n_{\text{tp}}$ is unreliable and can be inaccurate, with $w_{\text{eff}}$ varying greatly as a result of small statistics in the number of photon and coincidence counts as well as erroneous local minima.}
    \label{fig:TimeScaleHist}
\end{figure}

\begin{figure*}[tb!]
    \centering
    \makebox[\linewidth]{
    \subfloat{\includegraphics[width = 0.7\linewidth]{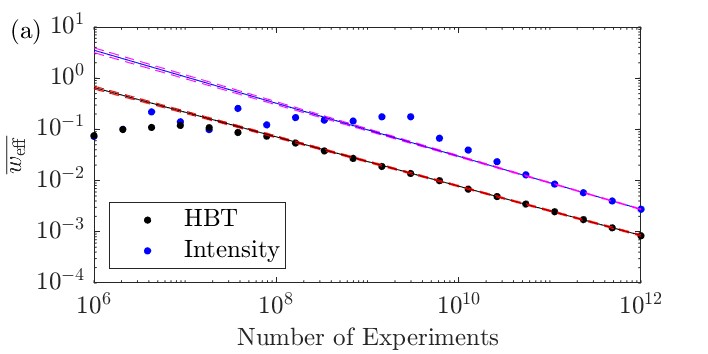}}
    \subfloat{\includegraphics[width = 0.7\linewidth]{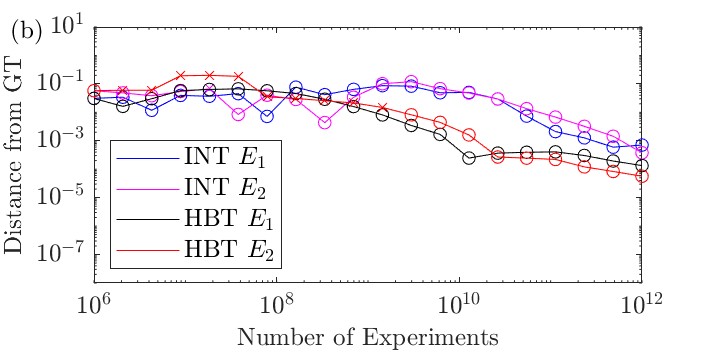}}}
    \makebox[\linewidth]{
    \subfloat{\includegraphics[width = 0.7\linewidth]{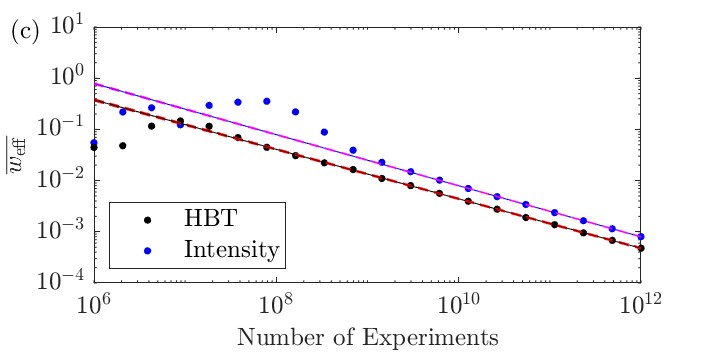}}
    \subfloat{\includegraphics[width = 0.7\linewidth]{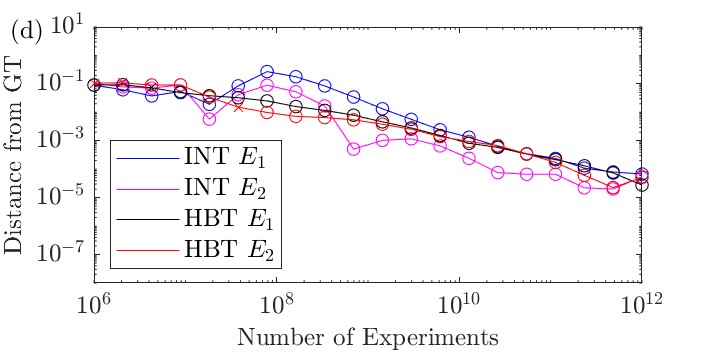}}}
    \makebox[\linewidth]{
    \subfloat{\includegraphics[width = 0.7\linewidth]{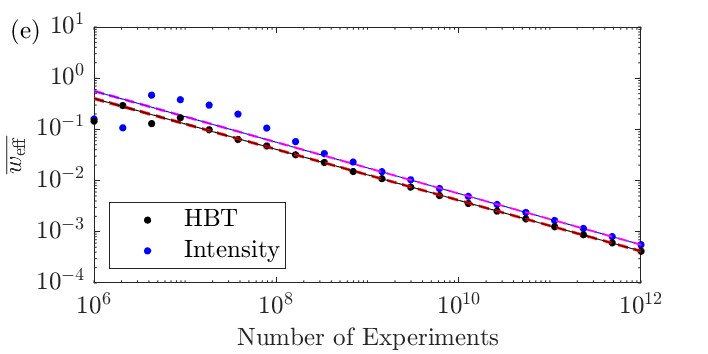}}
    \subfloat{\includegraphics[width = 0.7\linewidth]{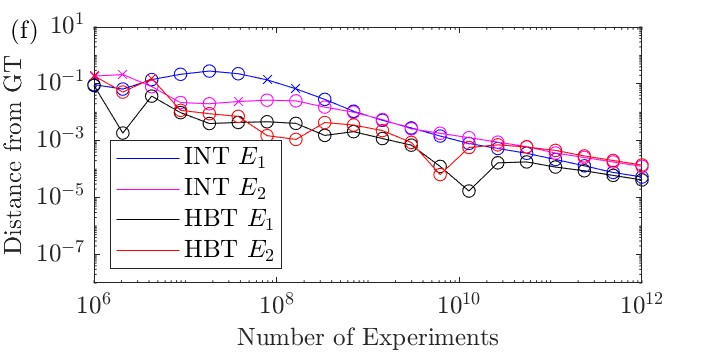}}}
    \caption{A comparison of $N$ intensity and $N$ HBT plus intensity detectors $\overline{w_{\text{eff}}}$ scaling results as a function of number of experiments, and distance of mean localisation position from ground truth (GT) for configurations with emitter spacings $0.11\sigma$ (a) \& (b), $0.21\sigma$ (c) \& (d), and $0.30\sigma$ (e) \& (f). We observe a turning point in the scaling behavior, $n_{\text{tp}}$, where the $\overline{w_{\text{eff}}}$ gradient converges on $1/\sqrt{n}$. We see that $n_{\text{tp}}$ occurs at lower $n$ when using HBT information as opposed to intensity alone. Additionally, $n_{\text{tp}}$ increases as emitter spacing decreases. The $1/\sqrt{n}$ scaling region is shown in (a), (c), \& (e) by the blue and black fit lines for intensity-only data and HBT data respectively. The hollow points in (b), (d), \& (f) show data where the GT lies within $w_\text{eff}$ polygon for each emitter, while crosses show where the GT does not. This demonstrates the potential inaccuracy of the fits when in the pre-$n_{\text{tp}}$ scaling regime.}
    \label{fig:TimePlots}
\end{figure*}

Figure~\ref{fig:TimeScaleHist} shows a histogram of  $\overline{w_{\text{eff}}}$ as a function of $n$ for 700 random emitter configurations. The percentage of the data that falls into 1 of 700 bins as $n$ increases is shown as a relative frequency for each value of $n$. Figure~\ref{fig:TimePlots} compares the $N$ HBT and $N$ intensity-only detector scaling results for three random cases with emitter spacings of $0.11\sigma$, $0.21\sigma$, and $0.33\sigma$. The results of these three cases are summarized in Table~\ref{TimeScaleTable}.

From these results, we observe that at some value $n$, the scaling begins to converge on a gradient $m$ of -0.5, corresponding to the expected $1/\sqrt{n}$ scaling on the logarithmic scale. We estimate the turning point where the gradient converges on $1/\sqrt{n}$ by fitting the data linearly in the region where a linear fit results in a gradient of -0.5, and ensure that the localisations are accurate, i.e., the emitter ground truths are contained within their $w_\text{eff}$. We term the estimated start point: $n_{\text{tp}}$. An alternative approach to estimating $n_{\text{tp}}$, as well as its relationship with emitter closeness, is studied in \cite{Jaret}. A similar metric to $n_{\text{tp}}$, denoted as $t_\text{knee}$, is used with the primary difference being that the estimation is done by interpolating the $w_\text{eff}$ gradient before and after $n_{\text{tp}}$. We find that the $n_{\text{tp}}$ approach leads to better turning point estimates for our widefield system.

\begin{figure}[tb!]
    \centering
    \makebox[\linewidth]{
    \subfloat{\includegraphics[width = 0.7\linewidth]{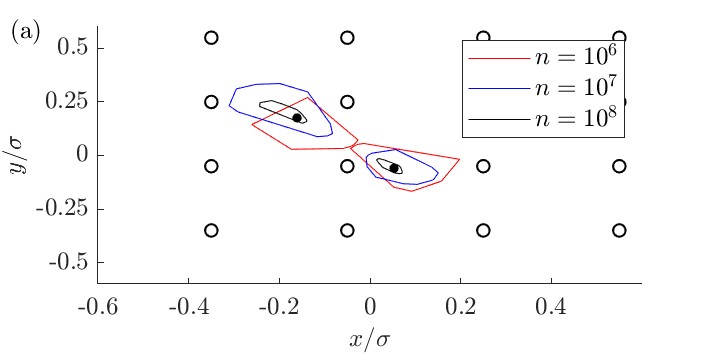}}
    \subfloat{\includegraphics[width = 0.7\linewidth]{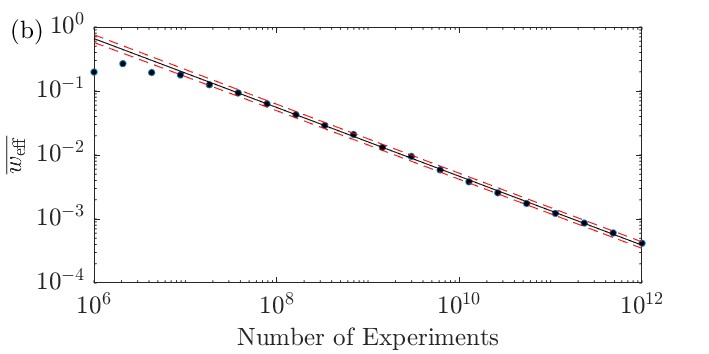}}}
    \caption{(a) Example $w_{\text{eff}}$ changing as $n$ increases for a configuration of emitter spacing $d = 0.32$ and brightness' $P_{\eta,1} = 0.4$ \& $P_{\eta,2} = 0.32$. (b) The full $n$ scaling data for the configuration. The $\overline{w_{\text{eff}}}$ at number of experiments $n = 10^{6}$ and $n = 10^{7}$ are prior to the $1/\sqrt{n}$ scaling regime, with only marginal improvement in mean $\overline{w_{\text{eff}}}$ of 0.20 to 0.18. There is significant improvement once $n_{\text{tp}}$ has been passed at $n = 10^{8}$, with a $\overline{w_{{\text{eff}}}}$ of 0.063.}
    \label{fig:TimeWeffPlot}
\end{figure}

In Figure~\ref{fig:TimeWeffPlot}, we show an example configuration with an emitter spacing of $d = 0.32\sigma$ and how $w_{\text{eff}}$ evolves over number of experiments $n = 10^{6}$, $n = 10^{7}$, and $n = 10^{8}$. For this configuration, $\overline{w_{\text{eff}}}$ changes from 0.18, to 0.16, and 0.056 for $n = 10^{6}$, $n = 10^{7}$, and $n = 10^{8}$, respectively. As $n = 10^{6}$ and $n = 10^{7}$ are before $n_{\text{tp}}$, there is marginal improvement in $w_{\text{eff}}$ until $n = 10^{8}$ where scaling becomes $1/\sqrt{n}$. Detector positions are marked as hollow circles. It is possible that for a low number of experiments that the solutions may fall in a local minima, resulting in an erroneous $w_{\text{eff}}$. 

We observe in our data that $n_{\text{tp}}$ increases as emitter spacing decreases. This is expected, as the $1/\sqrt{n}$ scaling regime begins once the emitters are individually resolved, which requires more photons when the emitters are closer.

When comparing the HBT to intensity-only results, we also observe that $n_{\text{tp}}$ is decreased when using HBT data compared to intensity-only data, with the difference being more significant for closer emitter spacings. As photon correlations are dependent on the product of $P_{1,k}$ and $P_{2,k}$, closer emitters result in an increase in correlated photon counts. This is additional data used in the localisation process, thereby reducing the $n$ required to localise the emitters.

The data in Figures~\ref{fig:TimePlots} and Table~\ref{TimeScaleTable}, show reductions in $n_{\text{tp}}$ when using HBT data of 1.13 $\times$ $10^{11}$ to 2.98 $\times$ $10^{9}$, 6.16 $\times$ $10^{9}$ to 3.36 $\times$ $10^{8}$, and 1.44 $\times$ $10^{9}$ to 3.79 $\times$ $10^{7}$ for cases (a), (b), and (c) in Figure~\ref{fig:TimePlots}, respectively. The dependency of the advantage gained when using HBT on emitter closeness will be of great importance when we compare systems of $N$ vs $N/2$ detector numbers in Section~\ref{Sect:HBTAdvantage}.

\begin{table}[tb!]
\caption{Summary of cases in Figure~\ref{fig:TimePlots}. gradient $m$ after the $\overline{w_{\text{eff}}}$ turning point, $n_{\text{tp}}$, and first data point in the $n_{\text{tp}}$ region are shown. We observe a lower $n_{\text{tp}}$ when using HBT plus intensity data, suggesting quicker emitter localisation than when using intensity alone. $n_{\text{tp}}$ increases with emitter closeness, as emitters in these cases are more difficult to resolve.}
\begin{center}
\makebox[\linewidth]{
\begin{tabular}{| c | c | c | p{2cm} | p{2cm} | p{2cm} | p{2cm} |}
\hline
\hline
Case & $d$ [$\sigma$] & $P_{\eta,1}$ \& $P_{\eta,2}$ & $m$ (HBT) & $m$ (INT) & $n_{\text{tp}}$ (HBT) & $n_{\text{tp}}$ (INT) \\
 \hline \hline
Fig.~\ref{fig:TimePlots}(a) & 0.11 & 0.4 \& 0.71 & -0.481 $\pm$ 0.006 & -0.52 $\pm$ 0.02 & 2.98 $\times$ $10^{9}$ & 1.13 $\times$ $10^{11}$ \\ \hline
Fig.~\ref{fig:TimePlots}(b) & 0.21 & 0.4 \& 0.328 & -0.483 $\pm$ 0.004 & -0.499 $\pm$ 0.003 & 3.36 $\times$ $10^{8}$ & 6.16 $\times$ $10^{9}$ \\ \hline
Fig.~\ref{fig:TimePlots}(c) & 0.33 & 0.4 \& 0.22 & -0.497 $\pm$ 0.004 & -0.500 $\pm$ 0.004 & 3.79 $\times$ $10^{7}$ & 1.44 $\times$ $10^{9}$ \\ \hline
\hline
\end{tabular}
}
\end{center}
\label{TimeScaleTable}
\end{table}

\newpage

\subsection{Localisation precision with respect to detector spacing for random emitter configurations}\label{sect:SpaceLoc}

As we increase the spacing of our detectors, the number of detectors, $N$, that we can fit within a defined area will decrease. Before we begin comparing $N$ vs $N/2$ $w_\text{eff}$ results for intensity-only vs HBT plus intensity, we determine the effect that this will have on $w_\text{eff}$.

\begin{figure}[tb!]
    \centering
    \makebox[\linewidth]{
    \subfloat{\includegraphics[width = 0.7\linewidth]{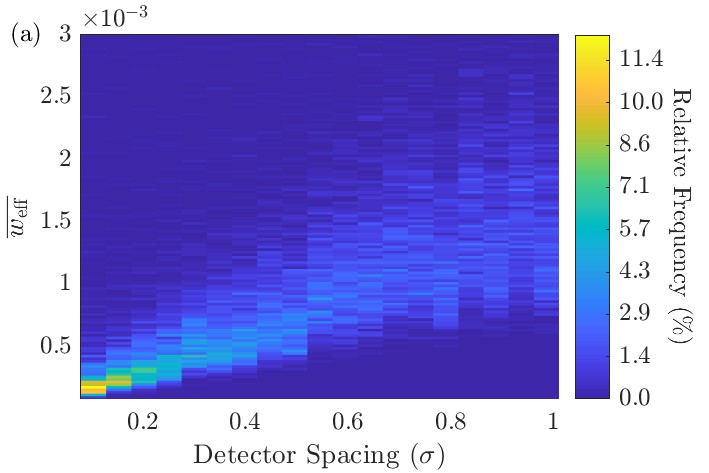}}
    \subfloat{\includegraphics[width = 0.7\linewidth]{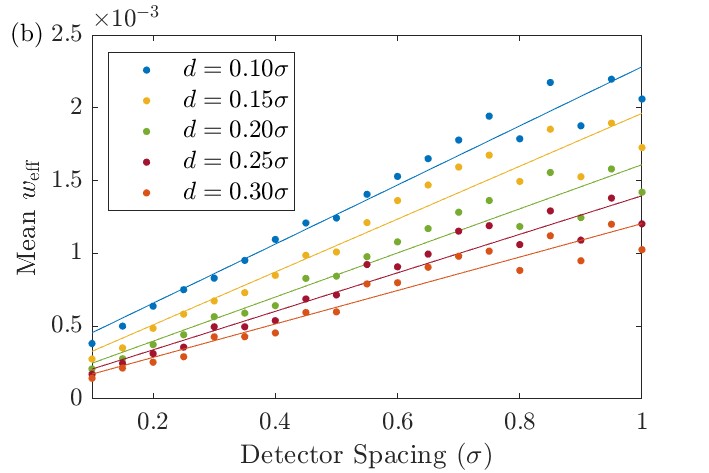}}}
    \caption{(a) $\overline{w_{\text{eff}}}$ (mean of two-emitter $w_{\text{eff}}$ values per case) changing as a function of detector spacing. We observe here a mostly linear relation, with the primary factor affecting scaling behavior being emitter spacing. A smaller emitter spacing results in a higher $\overline{w_{\text{eff}}}$ gradient. (b)  Mean $w_{\text{eff}}$ values of 20 different two-emitter configurations as a function of detector spacing. Emitters are spaced at distances from $0.10\sigma$ to $0.30\sigma$ in increments of $0.05\sigma$. We observe that for closer emitters, the rate at which $w_{\text{eff}}$ increases with detector spacing is higher.}
    \label{fig:SpacePlots}
\end{figure}

Figure~\ref{fig:SpacePlots}(a) shows how the mean $w_\text{eff}$ scales with detector spacing for 700 random emitter configurations for $10^{12}$ number of experiments. We choose a large number of experiments to ensure that the majority of the random configuration will be in the fully resolved regime, i.e., after $n_{\text{tp}}$. We note that changing the spacing of the detectors will influence the edges of the field of view, potentially resulting in a loss in photon detections near the boundaries. We account for this by increasing the bounds of the field of view to be very large ($-8\sigma$ to $8\sigma$ in the $x$ and $y$ axis). This leaves us with a buffer zone near the edge of the field of view, where detectors detect approximately zero photons. From Figure~\ref{fig:SpacePlots}(a) we observe a mostly linear relation, with $w_\text{eff}$ increasing as the detector spacing increases. As in section~\ref{sect:TimeLoc} the primary factor affecting $w_\text{eff}$ scaling behavior is emitter spacing, with a smaller spacing resulting in a larger $w_{\text{eff}}$. 

In Figure~\ref{fig:SpacePlots}(b), we show the $w_{\text{eff}}$ scaling for 5 different emitter spacings averaged over 20 random cases. The results are summarized in Table~\ref{SpaceScaleTable}. The data in Figure~\ref{fig:SpacePlots}(b) and Table~\ref{SpaceScaleTable} shows that $w_{\text{eff}}$ decreases as detector spacing decreases. This relationship appears roughly linear, with each plot being fitted by a gradient, $m$. Additionally, we see that as emitter spacing $d$ decreases, $w_{\text{eff}}$ increases, with $m$ also increasing. We see our largest $m$ of 0.0020 $\pm$ 0.0002 at $d$ = $0.10\sigma$ and smallest of 0.0011 $\pm$ 0.0002 at $d$ = $0.30\sigma$.

\begin{table*}[tb!]
\caption{Summary of results in Figure~\ref{fig:SpacePlots}, showing $w_{\text{eff}}$ as a function of detector spacing for increasing emitter spacing, $d$. Results are averaged over 20 emitter configurations with randomized positions and relative brightness. We observe an decreased gradient, $m$, as $d$ increases, meaning the localisation precision worsens more rapidly as detector spacing increases for closer emitters.}
\begin{center}
\makebox[\linewidth]{
\begin{tabular}{| c || p{2cm} | p{2cm} | p{2cm} | p{2cm} | p{2cm} |}
\hline
\hline
$d$ [$\sigma$] & 0.10 & 0.15 & 0.20 & 0.25 & 0.30 \\ \hline
$m$ (Gradient) & 0.0020 $\pm$ 0.0002 & 0.0018 $\pm$ 0.0002 & 0.0015 $\pm$ 0.0002 & 0.0013 $\pm$ 0.0002 & 0.0011 $\pm$ 0.0002 \\ \hline
\hline
\end{tabular}
}
\end{center}
\label{SpaceScaleTable}
\end{table*}

\subsection{Comparison of intensity and HBT for $N$ and $N/2$ detectors}\label{Sect:HBTAdvantage}

From the previous sections we see that the use of HBT plus intensity can improve localisation precision, and that localisation is improved with more detectors, closer, within a fixed area. However, when the total number of SPAD detectors is limited, we must consider that $N$ detectors can only perform HBT measurements at $N/2$ locations. In order to justify the use of HBT, the $w_{\text{eff}}$ for $N/2$ HBT detectors, would need to be equal to or less than the $w_{\text{eff}}$ for $N$ intensity detectors.

To compare $N$ intensity vs $N/2$ HBT detectors, we take the ratio of the intensity mean $w_{\text{eff}}$ over the mean HBT $w_{\text{eff}}$. We term this metric the HBT advantage, where a value greater than 1 indicates an improvement in localisation precision when using HBT by a factor of that value.

As our array configurations are square grids, it would typically be impossible to make a direct comparison of $N$ and $N/2$ detectors in a square configuration. Therefore, we approximate the result for $N$ and $N/2$ by calculating how many detectors would fit in a square with an area of $\sigma$ using a set detector spacing, $D_{s}$. We do this using:

\begin{equation}
    D_S = 1/(\sqrt{N}-1).
\label{eq:NtoDS}
\end{equation}

\begin{figure}[tb!]
    \centering
    \makebox[\linewidth]{
    \subfloat{\includegraphics[width = 0.7\linewidth]{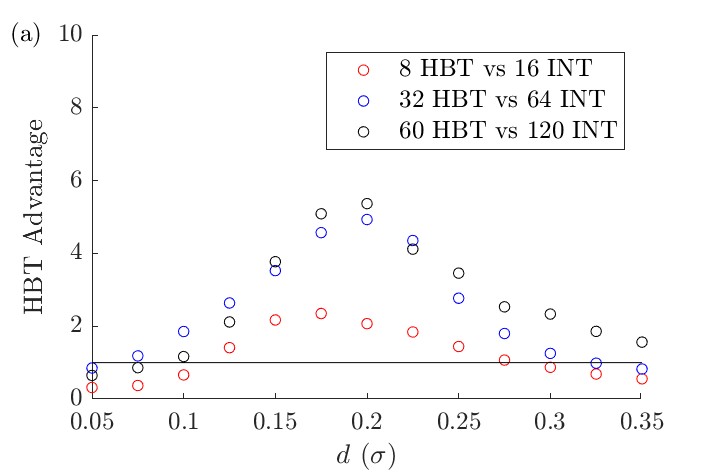}}
    \subfloat{\includegraphics[width = 0.7\linewidth]{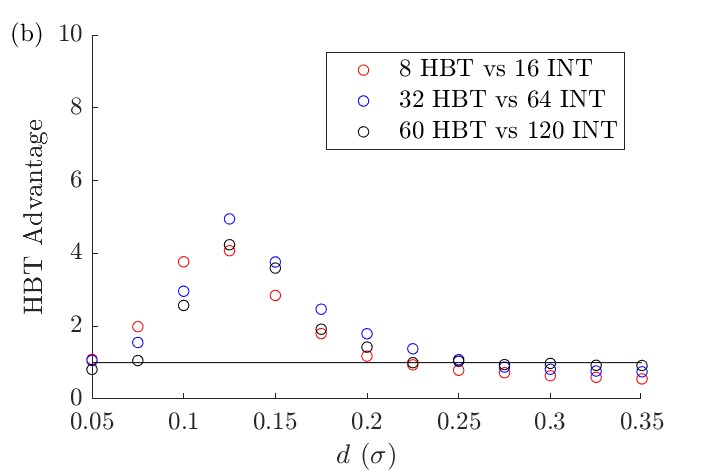}}}
    \includegraphics[width = 0.7\linewidth]{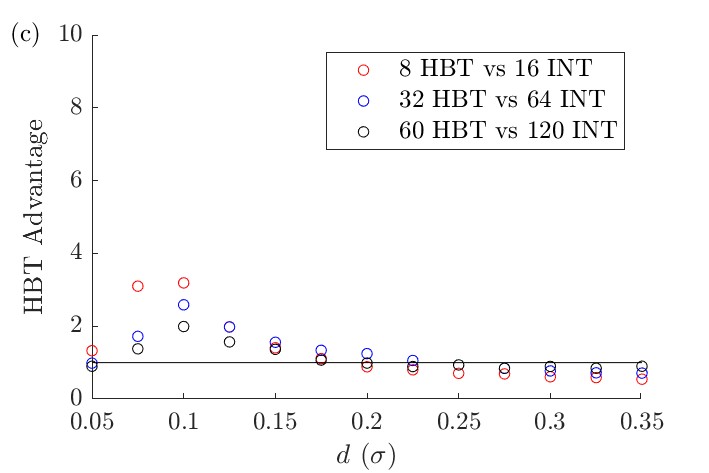}
    \caption{A comparison of HBT advantage for three different detector array configurations for number of expeirments $n = 10^{8}$ (a), $n = 10^{9}$ (b), and $n = 10^{10}$ (c). HBT advantage is calculated by taking the ratio of intensity-only $\overline{w_{\text{eff}}}$ over HBT $\overline{w_{\text{eff}}}$. The black line across HBT advantage = 1 signifies the threshold where HBT outperforms intensity-only. The region to the left of the peak in each figure are the emitter spacings that are unresolved within the given number of experiments.}
    \label{fig:HBT_Adv}
\end{figure}

\begin{figure}[tb!]
    \centering
    \makebox[\linewidth]{
    \subfloat{\includegraphics[width = 0.7\linewidth]{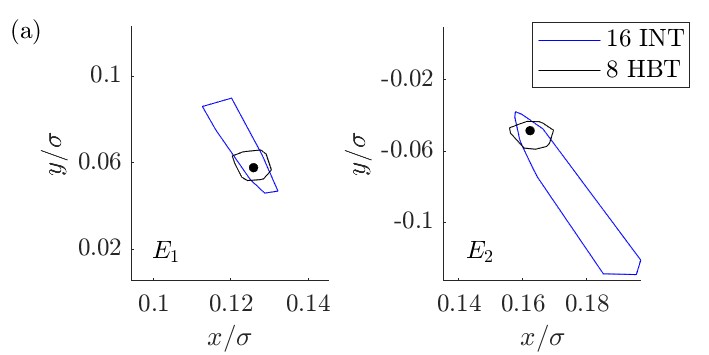}}
    \subfloat{\includegraphics[width = 0.7\linewidth]{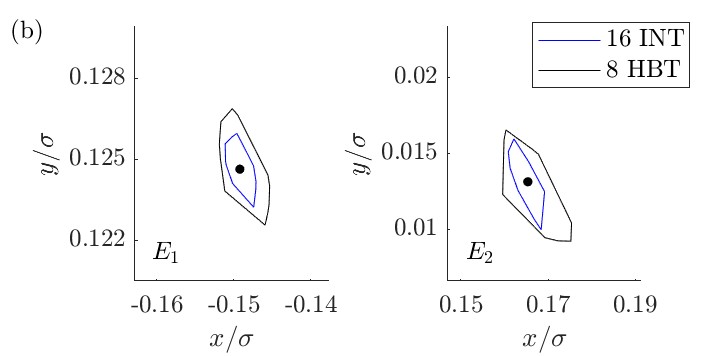}}}
    \makebox[\linewidth]{
    \subfloat{\includegraphics[width = 0.7\linewidth]{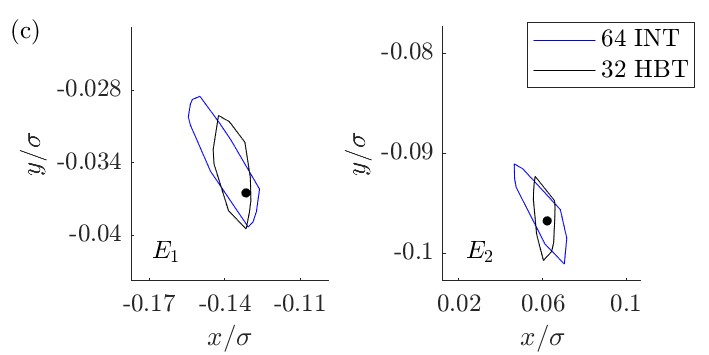}}
    \subfloat{\includegraphics[width = 0.7\linewidth]{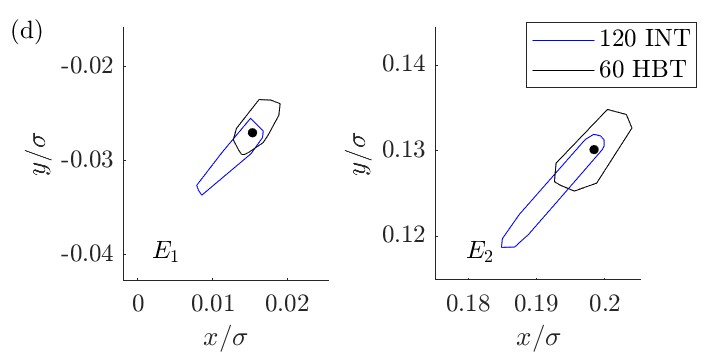}}}
    \caption{Example $w_{\text{eff}}$ comparisons for $N$ intensity-only vs $N/2$ HBT plus intensity detectors. (a) shows a case with an emitter spacing of $0.11\sigma$ and brightness' $P_{\eta,1} = 0.4$ \& $P_{\eta,2} = 0.284$ at $n = 10^{10}$. (b) shows a case with an emitter spacing of $0.33\sigma$ and brightness' $P_{\eta,1} = 0.4$ \& $P_{\eta,2} = 0.22$ at $n = 10^{10}$. (c) shows a case with an emitter spacing of $0.20\sigma$ and brightness' $P_{\eta,1} = 0.4$ \& $P_{\eta,2} = 0.344$ at $n = 10^{9}$. (d) shows a case with an emitter spacing of $0.24\sigma$ and brightness' $P_{\eta,1} = 0.4$ \& $P_{\eta,2} = 0.296$ at $n = 10^{9}$. Note that for cases (a) and (c), the HBT $\overline{w_{\text{eff}}}$ is smaller, resulting in a HBT advantage of 2.2 and 1.2 respectively. Conversely, cases (b) and (d) have a smaller intensity-only $\overline{w_{\text{eff}}}$, with HBT advantage scores of 0.60 and 0.96 respectively. Whether $N$ intensity-only or $N/2$ HBT detectors are superior for a given number of experiments is tied to the closeness of the emitters.}
    \label{fig:HBT_Adv_Weffs}
\end{figure}

In Figure~\ref{fig:HBT_Adv} we show the HBT advantage as a function of emitter spacing, $d$, averaged over 20 emitter configurations for each $d$ for number of experiments: $10^{8}$, $10^{9}$, and $10^{10}$. In Figure~\ref{fig:HBT_Adv_Weffs}, we show example $w_{\text{eff}}$ plots for 4 random emitter configurations. The information and results of these cases is summarized in Table~\ref{Adv_Weff_Table}.

 We observe a peak for each plot in Figure~\ref{fig:HBT_Adv}, showing for what emitter spacings the HBT advantage is most significant. As number of experiments decreases, the position of the peak increases from an emitter spacing of $0.2\sigma$, to $0.125\sigma$, and to $0.1\sigma$ number of experiments $10^{8}$, $10^{9}$, and $10^{10}$, respectively. The data to the left of the peak corresponds to emitter configurations that have not been localised for the given $n$. The peak is due to the $n_{\text{tp}}$ properties discussed in Section \ref{sect:TimeLoc}. The $\overline{w_{\text{eff}}}$ values when using HBT data are significantly lower in the regime when the emitters are localised on the basis of HBT, but not yet with intensity-only. For closer emitter spacing, there will be an increased probability of correlated photon detections, and hence an improved localisation precision when using $N/2$ HBT detectors relative to $N$ intensity-only detectors.
 
\begin{figure}[tb!]
    \centering
    \makebox[\linewidth]{
    \subfloat{\includegraphics[width = 0.7\linewidth]{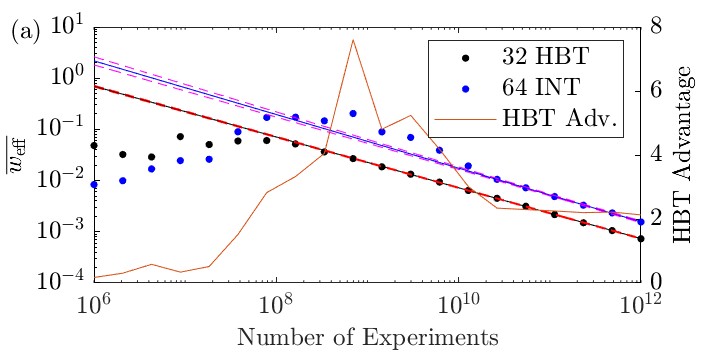}}
    \subfloat{\includegraphics[width = 0.7\linewidth]{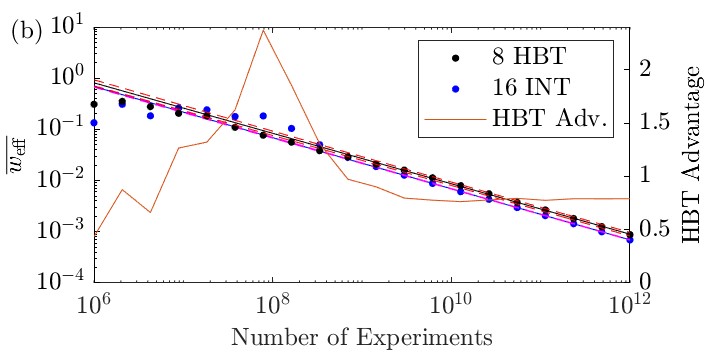}}}
\caption{$n$ scaling plots comparing $\overline{w_{\text{eff}}}$ for $N$ intensity-only detectors and $N/2$ HBT plus intensity detectors (left axis). The HBT advantage as a function of $n$ is also included (right axis). (a) shows the scaling and HBT advantage results for a case with an emitter spacing of $0.11\sigma$ and brightness' $P_{\eta,1} = 0.4$ \& $P_{\eta,2} = 0.264$. (b) shows the scaling and HBT advantage results for a case with an emitter spacing of $0.26\sigma$ and brightness' $P_{\eta,1} = 0.4$ \& $P_{\eta,2} = 0.256$. Due to $n_{\text{tp}}$ occurring earlier for the HBT $w_{\text{eff}}$, there is a region where the intensity $\overline{w_{\text{eff}}}$ is noticeably higher. The peak of (b) is smaller as the emitters are further appart, thereby lowering the number of correlated photons.}
\label{fig:HBT_Adv_Time}
\end{figure}

In Figure~\ref{fig:HBT_Adv_Time}, we see how $n_{\text{tp}}$ occurs earlier for the HBT $\overline{w_{\text{eff}}}$ compared to the intensity-only $\overline{w_{\text{eff}}}$ for the shown cases. Figure~\ref{fig:HBT_Adv_Time}(a) shows a case with an emitter spacing of $0.11\sigma$ and brightness' $P_{\eta,1} = 0.4$ \& $P_{\eta,2} = 0.264$. Here, the HBT $n_{\text{tp}}$ occurs at $n = 6.95 \times 10^8$, while the intensity $n_{\text{tp}}$ occurs at $n = 4.56 \times 10^{10}$. This results in a region where the intensity $\overline{w_{\text{eff}}}$ is noticeably higher than the HBT $\overline{w_{\text{eff}}}$. The HBT advantage as a function of $n$, which is also shown, peaks in this region before reducing in value at higher $n$. This behavior is in agreement with the results seen in Figure~\ref{fig:HBT_Adv} where, once the emitters are resolved using HBT, the HBT advantage decreases an $n$ increases due to the emitters becoming resolved by intensity-only measurements. A similar feature can be seen in Figure~\ref{fig:HBT_Adv_Time}(b) which shows a case with an emitter spacing of $0.26\sigma$ and brightness' $P_{\eta,1} = 0.4$ \& $P_{\eta,2} = 0.256$. Notably, due to the emitter distance being greater, the peak here is smaller and, once $n_{\text{tp}}$ is reached for the intensity $\overline{w_{\text{eff}}}$, the intensity $\overline{w_{\text{eff}}}$ outperforms the HBT $\overline{w_{\text{eff}}}$. This is also reflected by the HBT advantage plot becoming less than 1. 

Comparing our Figure~\ref{fig:HBT_Adv_Time} plots to the Figure~\ref{fig:HBT_Adv} plots, we can observe how and why the HBT advantage peak moves towards smaller emitter spacings as number of experiments increases. At larger $n$ values, emitters with spacings such as $0.2\sigma$ are able to be resolved with intensity-only data. In such cases, the increased photon detections gained from having $N$ detectors are more significant than the photon correlations measured by $N/2$ detectors, resulting in a HBT advantage of less than 1. HBT advantage remains greater than 1 at higher $n$ for very small emitter spacings as the increased number of correlated photons allows for smaller $\overline{w_{\text{eff}}}$. As stated, this requires that the emitters are resolved using HBT.

We show comparisons of $N/2$ HBT $w_{\text{eff}}$ and $N$ intensity $w_{\text{eff}}$ for 4 different emitter configurations in Figure~\ref{fig:HBT_Adv_Weffs}, with the HBT advantage results summarized in Table~\ref{Adv_Weff_Table}. Cases where intensity-only outperforms HBT can be observed in Figs.~\ref{fig:HBT_Adv_Weffs} (b) \& (d), where the emitter spacings of $0.33\sigma$ and $0.24\sigma$, respectively, are sufficiently large and number of experiments sufficiently high that $N$ intensity detectors outperforms HBT, with HBT advantage scores of 0.60 and 0.96.

It should also be noted that for lower values of $n$, such as less than $10^{8}$, the number of detected counts is reduced and thus, the probability of coincident photon events occurring enough times to be used reliably for localisation is very low, resulting in very noisy data or results indistinguishable from using $N/2$ intensity-only detectors. This effect can begin to be observed in Fig.~\ref{fig:HBT_Adv}(a) where the 8 VS 16 detector HBT advantage peak at $d = 0.2\sigma$ is lower than the 32 VS 64 detector, and 60 VS 120 detector peaks by a factor of 2.38 and 2.95, respectively. Conversely, this behavior is not as significant when $n$ is greater, as seen in Fig.~\ref{fig:HBT_Adv}(b) where at $d = 0.125\sigma$, the 8 VS 16 detector HBT advantage peak is lower than the 32 VS 64 detector, and 60 VS 120 detector peaks by a factor of 1.04 and 1.21, respectively. Increasing the number of correlations detected at low values of $n$ would require a system with decreased optical losses, $\eta$, or emitters with an increased intrinsic brightness, $P_{B,i}$.

\begin{table*}[tb!]
\caption{Summarized results of Figure~\ref{fig:HBT_Adv_Weffs} where the $\overline{w_{\text{eff}}}$ of $N$ intensity-only and $N/2$ HBT detectors are compared. The HBT advantage is the ratio of the intensity-only $\overline{w_{\text{eff}}}$ and HBT $\overline{w_{\text{eff}}}$. A value of greater than 1 indicates that the $N/2$ HBT detectors provide superior localisation precision.}
\begin{center}
\makebox[\linewidth]{
\begin{tabular}{| c | c | c | c | c | l | l | l |}
\hline
\hline
\makecell[c]{Case \\Fig.} & $d$ [$\sigma$] & $P_{\eta,1}$ \& $P_{\eta,2}$ & \makecell[c]{Number of \\ Experiments} & Detector Numbers & $\overline{w_{\text{eff}}}$ (INT) & $\overline{w_{\text{eff}}}$ (HBT) & \makecell[c]{HBT \\ Advantage} \\
 \hline \hline
\ref{fig:HBT_Adv_Weffs}(a) & 0.11 & 0.4 \& 0.284 & $10^{10}$ & 16 INT vs 8 HBT & 0.0224 & 0.0118 & 1.9 \\ \hline
\ref{fig:HBT_Adv_Weffs}(b) & 0.33 & 0.4 \& 0.22 & $10^{10}$ & 16 INT vs 8 HBT & 0.0038 & 0.0061 & 0.62 \\ \hline
\ref{fig:HBT_Adv_Weffs}(c) & 0.20 & 0.4 \& 0.344 & $10^{9}$ & 64 INT vs 32 HBT & 0.0099 & 0.0086 & 1.2 \\ \hline
\ref{fig:HBT_Adv_Weffs}(d) & 0.24 & 0.4 \& 0.296 & $10^{9}$ & 120 INT vs 60 HBT & 0.0054 & 0.0061 & 0.89 \\ \hline
\hline
\end{tabular}
}
\end{center}
\label{Adv_Weff_Table}
\end{table*}

\subsection{Advantage of quantum correlations as a function of photon detection efficiency}

So far, we have observed an increase in HBT advantage for configurations of closer emitters. We expect these cases to be difficult to resolve on the basis of intensity alone, requiring additional (quantum) information to be able to resolve the emitters with fewer photons. Our results suggest that a primary factor leading to the increase in HBT advantage is the increase in coincident counts as emitters are closer. As the ratio of coincident counts to photon counts increases, so too does the HBT advantage.

To investigate this effect, we consider some configurations of different emitter spacings, number of experiments, and photon detection efficiency, $1-\eta$, where $\eta$ is the optical losses of the system as described in Section~\ref{sect:2A}. We keep the intrinsic brightness of the emitters the same ($P_{B,1} = P_{B,2}$), but increase $1-\eta$ to increase photon detection rates. This will also increase the ratio of coincident counts to photon counts, which we expect would result in a increase in HBT advantage for each configuration.

\begin{figure}[tb!]
    \centering
    \includegraphics[width = 0.7\linewidth]{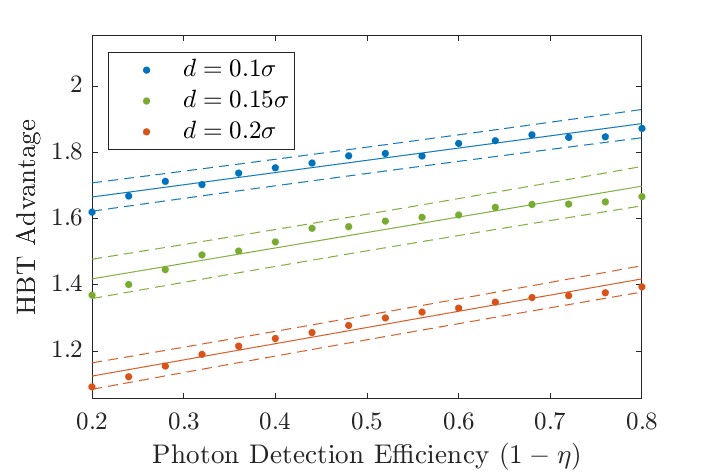}
    \caption{HBT Advantage for emitter spacings $d =0.1\sigma$, $d =0.15\sigma$, and $d =0.2\sigma$, as a function of photon detection efficiency. Results are averaged over 20 random emitter configurations. We observe that an increase in detector efficiency leads to an increased number of photon counts, thereby increasing the total number of coincidences. This increases the HBT advantage, and highlights that the advantage of QCM increases as the number of counts increases, albeit by a factor that depends on the exact microscopic properties of the system.}
    \label{fig:PB_HBTAdv}
\end{figure}

Figure~\ref{fig:PB_HBTAdv} shows HBT advantage as a function of $1-\eta$, for different emitter spacings. Results are averaged over 20 different configurations. Number of experiments is set to $10^{12}$ pulses.

We obtain a gradient of 0.37 $\pm$ 0.06, 0.47 $\pm$ 0.08, and 0.49 $\pm$ 0.05 for emitter spacings $0.1\sigma$, $0.15\sigma$, and $0.2\sigma$, respectively. The increasing gradient shows that the HBT advantage is more significant for systems with greater photon detection efficiency.

\section{Conclusion}

When comparing $N$ intensity-only detector measurements and $N$  HBT plus intensity detector measurements, our results show that the use of HBT measurements will result in an improvement in localisation precision, with the factor of improvement being dependent on the spacing of the emitters.

Our $w_{\text{eff}}$ scaling as a function of number of experiments shows that the HBT data reaches $n_{\text{tp}}$, the start of the resolved emitter regime, faster than using intensity-only. We observe, however, that for a low number of experiments, that the localisation results may not entirely be accurate relative to the ground-truth emitter positions. Extra consideration may be necessary when examining cases where the HBT or intensity data appears the reach the $n_{\text{tp}}$ regime at lower number of experiments to ensure that the localisation is actually accurate.

We also observe a dependency of emitter spacing for $w_{\text{eff}}$ scaling as a function of detector spacing, with closer emitters scaling more poorly as detector separation increases. We expect that $w_{\text{eff}}$ would be poorer when the detector spacing is far relative to the emitter spacing due to a narrower PSF.

When considering $N$ intensity-only detectors and $N/2$ HBT plus intensity detectors, our results find that there is an ideal regime when an advantage is gained from using $N/2$ HBT detectors when the emitters are close. The size of this regime increases as number of experiments decreases, as the emitters are localised on the basis of HBT, but not using intensity-only. The HBT advantage increasing as emitter spacing decreases requires that the emitters are resolved using HBT, else the advantage disappears. The advantage is also not present for larger emitter spacings for higher numbers of experiments.

For the HBT advantage to be more significant and persist for larger spacings at higher measurement times would require the system to have an increased photon detection efficiency so that the ratio of correlated counts and photon counts are increased. Our current hypothetical system requires numbers of experiments in the order of $10^8$ to accurately localise emitters. Assuming a pulse cycle of 100 ns, this would require a real time of 10 s to localise emitters. There is potential for improvement if a more effective system is considered or when considering emitters with a lifetime that would result in more photon emissions for a shorter time frame.

\section{Acknowledgments}

The authors acknowledge the assistance from the RMIT Center of Excellence for Nanoscale BioPhotonics. We also acknowledge the use of the RMIT AWS Cloud Supercomputing Hub resources. This work is funded by the Air Force Office of Scientific Research (FA9550-20-1-0276).



\bibliography{References}

\end{document}